\documentstyle[aps,multicol,epsfig]{revtex}

\draft

\begin{document}

\title{\bf Optomechanical scheme for the detection of 
weak impulsive forces}
\author{David Vitali$^1$, Stefano Mancini$^{1,2}$, and Paolo Tombesi$^1$}
\address{
$^1$INFM, Dipartimento di Matematica e Fisica,
Universit\`a di Camerino,
I-62032 Camerino, Italy \\
$^2$INFM, Dipartimento di Fisica,
Universit\`a di Milano,
Via Celoria 16, I-20133 Milano, Italy}

\date{Received: \today}

\maketitle

\begin{abstract}
We show that a cooling
scheme and an appropriate quantum nonstationary strategy 
can be used to improve the signal to noise ratio 
for the optomechanical detection of weak impulsive forces.

\end{abstract}


\begin{multicols}{2}

A mechanical oscillator coupled to an optical mode
by the radiation pressure provides a sensitive device able to
detect very weak forces. Relevant examples are interferometers
for the detection of gravitational waves \cite{GRAV} and atomic
force microscopes \cite{AFM}.
Up to now, the major limitation to the implementation
of sensitive optical measurements
is given by thermal noise \cite{HADJAR}.
It has been proposed in Ref.~\cite{MVTPRL} to reduce thermal noise
by means of a feedback loop based on homodyning the light 
reflected by the oscillator, playing the role of a cavity mirror.
The proposed scheme is a sort of continuous version of the 
stochastic cooling technique used in accelerators \cite{acc}, 
because the homodyne 
measurement provides a continuous monitoring of the oscillator's
position, and the feedback continuously
``kicks'' the mirror in order to put it in its
equilibrium position. This proposal has been experimentally realized in
Ref.~\cite{HEIPRL}, using the ``cold damping'' technique \cite{coldd}, 
which is 
physically analogous to that proposed in Ref.~\cite{MVTPRL} and which
amounts to applying a viscous feedback force to the oscillating mirror.

Both the ``stochastic cooling'' scheme of Ref.~\cite{MVTPRL} and the cold 
damping scheme of Ref.~\cite{HEIPRL} cool the mirror by overdamping it, 
thereby strongly decreasing its mechanical susceptibility at resonance.
As a consequence, the oscillator does not resonantly respond to the 
thermal noise, yielding
in this way an almost complete suppression of the resonance peak in the 
noise power spectrum, which is equivalent to cooling.
However, the two feedback schemes cannot be directly applied 
to improve the detection
of weak forces. In fact
the strong reduction of the mechanical susceptibility at resonance means that
the mirror does not respond not only to the noise but also to the signal, and 
we shall see that the signal to noise ratio (SNR) of the device
in stationary conditions is actually never improved.
Despite that, here we show how it is possible to
design a {\em nonstationary} strategy
able to significantly increase the SNR for the detection
of {\em impulsive} classical forces acting on the oscillator.
This may be of crucial importance in the
field of metrology \cite{ROUK}, as well as for
the detection of gravitational waves \cite{GRAV}.
We use a quantum treatment, allowing us to show why a 
classical approach provides an incomplete description
of the optomechanical scheme.

Let us consider a simplified system
with a single mechanical mode, representing the movable
mirror (with mass $m$ and frequency 
$\omega_{m}$) of a coherently driven optical cavity. 
The optomechanical coupling between the mirror and
the cavity field is realized by the radiation pressure.
In the adiabatic limit in which the mirror frequency is much smaller 
than the cavity free spectral range $c/2L$ ($L$ is the cavity length)
\cite{LAW}, one can focus on one cavity mode only (with 
annihilation operator $b$, frequency $\omega_c$ and cavity decay rate
$\gamma_c$)
because photon scattering into other modes can be neglected.
This adiabatic regime implies $\omega_{m} \ll \omega_{c}$,
and therefore the generation of photons due to the Casimir effect,
retardation and Doppler effects are completely negligible.
The cavity mode is driven by a laser field with input power $\wp$
and frequency $\omega_0 \sim \omega_c$.
The dynamics of the system
can be described by the following set of coupled quantum 
Langevin equations (QLEs)
(in the interaction picture with respect to 
$\hbar \omega_{0}b^{\dagger} b$) 
\begin{mathletters}
\label{QLENL}
\begin{eqnarray}
\dot{Q}(t) &=& \omega_m P(t), 
\label{QLENL1}\\
\dot{P}(t) &=& -\omega_{m} Q(t) -  
 {\gamma_m} P(t)   
 +G b^{\dagger}(t) b(t) \nonumber \\
 &+& {\cal W}(t) +f(t),
\label{QLENL2}\\
\dot{b}(t) &=& \left(-i \omega_{c} + i \omega_{0} -
\frac{\gamma_{c}}{2}\right) b(t) + 2i 
G Q(t) b(t) \nonumber \\
&+& E +\sqrt{\gamma_{c}}b_{in}(t), 
\label{QLENL3}
\end{eqnarray}
\end{mathletters}
where $Q$ and $P$ are the dimensionless position and momentum
operator of the movable mirror, $\gamma_m$ is the mechanical damping rate,
$G=(\omega_c/L)\sqrt{\hbar/2m\omega_m}$ is the coupling constant,
$f(t)$ is the classical force to be detected,
and $E=\sqrt{\wp\gamma_{c}/\hbar \omega_{0}}$.
The noise terms in the QLEs are given by
the usual input noise operator $b_{in}(t)$ \cite{milwal}, 
associated with the vacuum fluctuations of the continuum of 
electromagnetic modes outside the cavity, and by the
random force ${\cal W}(t)$ describing
the Brownian motion of the mirror caused by the 
coupling with other internal and external modes at the 
equilibrium temperature $T$.
The optical input noise correlation function is
$\langle b_{in}(t)b_{in}^{\dagger}(t') \rangle = \delta(t-t') $
\cite{milwal}, while that of the quantum Langevin force ${\cal W}(t)$
is given by \cite{GAR,VICTOR}
$
\langle {\cal W}(t) {\cal W}(t^\prime) \rangle=
\left(\gamma_m/2\pi \omega_m\right)\left[{\cal F}_{r}(t-t') - i 
{\cal F}_{i}(t-t') \right]$
where $
{\cal F}_{r}(t)=\int_{0}^{\varpi} d\omega
  \omega \cos(\omega t) \coth\left(\hbar\omega/2 k_B T \right)$,
${\cal F}_{i}(t)= \int_{0}^{\varpi} d\omega
  \omega \sin(\omega t) $
with $\varpi$ the frequency cutoff of the reservoir
spectrum. 
The QLEs (\ref{QLENL}), supplemented with the above correlation
functions, provide an {\em exact} description
of the system dynamics, valid at all temperatures
\cite{VICTOR}. 

In standard interferometric applications, the driving field is
very intense. Under this condition the system is characterized by a
semiclassical steady state with 
the internal cavity mode in a coherent
state $|\beta\rangle $, and a new equilibrium position for the 
mirror, displaced by $G| \beta |^{2}/\omega_m$.
Then the dynamics is well described 
by linearizing the QLEs 
(\ref{QLENL}) around the steady state, and if we 
rename with $Q(t)$ and $b(t)$ the operators describing the quantum 
fluctuations around the classical steady state, one gets
\begin{mathletters}
\label{QLEL}
\begin{eqnarray}
\dot{Q}(t) &=& \omega_m P(t) \,, 
\label{QLEL1}\\
\dot{P}(t) &=& -\omega_{m} Q(t) -  
  {\gamma_m} P(t)   +G\beta
  \left[ b(t) + 
  b^{\dagger}(t) \right]\nonumber \\
  & + & {\cal W}(t) +f(t), \label{QLEL2} \\ 
\dot{b}(t) &=&  \left(-\frac{\gamma_{c}}{2} - i \Delta\right)b(t)
+2i G \beta Q(t)+ \sqrt{\gamma_{c}} b_{in}(t) \,,
\label{QLEL3}
\end{eqnarray}
\end{mathletters} 
where we have chosen the phase of the cavity mode field so that
$\beta$ is real and $
	\Delta = \omega_{c} - \omega_{0} - G^2 \beta^{2}/\omega_m $	
is the cavity mode detuning. 
We shall consider from now on $\Delta=0$, 
which can
always be achieved by appropriately adjusting $\omega_{0}$. 
In this case the dynamics becomes 
simpler, because only the phase quadrature 
$Y(t)=i\left[b^{\dagger}(t)-b(t)\right]/2$ is 
affected by the mirror position fluctuations $Q(t)$,
while the amplitude field quadrature $X(t)=[b(t)+b^{\dagger}(t)]/2$ 
is not. 
The large cavity bandwidth limit $\gamma_{c} \gg G\beta $
is commonly considered, and in this case the
cavity mode dynamics adiabatically follows that of the mirror
position and it can be eliminated, i.e., 
\begin{equation}
Y(t) \simeq \frac{4G \beta}{\gamma_{c}}Q(t) 
+\frac{Y_{in}(t)}{\sqrt{\gamma_{c}}},
\label{adiab}
\end{equation}
where $Y_{in}(t)=i\left[b_{in}^{\dagger}(t)-b_{in}(t)\right]$.
In this limit, the movable mirror can be used as a ponderomotive meter 
to detect a weak force $f(t)$ acting on it \cite{BK},
which will be proportional to the displacement from the equilibrium 
position $Q(t)$.
The measured quantity is the output homodyne photocurrent,
$Y_{out}(t)=2\sqrt{\gamma_{c}}\eta 
Y(t)-\sqrt{\eta}Y_{in}^{\eta}(t)$ \cite{milwal},
where $\eta $ is the detection efficiency, and $Y_{in}^{\eta}(t)$
is a generalized input noise, coinciding with the
input noise $Y_{in}(t)$ in the case of perfect detection $\eta 
=1$, and taking into account the additional noise due to the 
inefficient detection in the general case $\eta < 1$ \cite{delay}.
This generalized input noise can be written as 
$Y_{in}^{\eta}(t)=i\left[b_{\eta}^{\dagger}(t)-b_{\eta}(t)\right]$, 
with $\langle b_{\eta}(t)b_{\eta}^{\dagger}(t')
\rangle = \delta(t-t')$, and it is correlated with the input noise $b_{in}(t)$
according to $\langle b_{in}(t)b_{\eta}^{\dagger}(t')
\rangle = \langle b_{\eta}(t)b_{in}^{\dagger}(t')
\rangle = \sqrt{\eta}\delta(t-t')$ \cite{delay}.

The output of the homodyne measurement may be used to devise a
phase-sensitive feedback loop to control the dynamics of the mirror.
The effect of the feedback loop has been 
described using quantum trajectory theory \cite{howmil} and the master 
equation formalism in Ref.~\cite{MVTPRL}, and a classical 
description neglecting all quantum fluctuations
in Ref.~\cite{HEIPRL}. Here we shall use a more general 
description of feedback based on QLEs for
Heisenberg operators, first developed in Ref.~\cite{howa} and 
generalized to the non-ideal detection case in Ref.~\cite{delay}.
In particular we shall give the first quantum description
of the cold damping scheme \cite{coldd}. 
The adoption of a fully quantum treatment is justified
by the fact that, as we shall see, in the presence of feedback
the radiation quantum noise has important effects, especially at low
temperatures.

In the proposal of Ref.~\cite{MVTPRL}, feedback induces position 
shifts controlled by the output homodyne photocurrent $Y_{out}(t)$.
This is described by an additional term in the QLE 
for a generic operator ${\cal O}(t)$ given by \cite{delay}
\begin{equation}\label{DOSTRA}
\dot{{\cal O}}_{fb}(t)=
i\frac{\sqrt{\gamma_{c}}}{\eta}
Y_{out}(t-\tau)
\left[g_{sc}P(t),{\cal O}(t)\right]\,,
\end{equation}
where $\tau$ is the feedback loop delay time, and $g_{sc}$ is the feedback 
gain. The feedback delay time is always much smaller than the typical 
timescale of the mirror dynamics and it can be neglected. After the 
adiabatic elimination of the cavity mode, the mirror QLEs become
\begin{mathletters}
\label{QLEFSCAD}
\begin{eqnarray}
\dot{Q}(t) &=& \omega_m P(t) 
+4G\beta g_{sc} Q(t) \nonumber \\
&-&\frac{g_{sc}}{2}\sqrt{\frac{\gamma_{c}}{\eta}}
Y_{in}^{\eta}(t)+g_{sc}\sqrt{\gamma_{c}}Y_{in}(t), 
\label{QLEFSCAD1}\\
\dot{P}(t) &=& -\omega_{m} Q(t) -  
  {\gamma_m} P(t)   +\frac{2G\beta}{\sqrt{\gamma_{c}}}
  X_{in}(t)  \nonumber \\
  &+& {\cal W}(t) +f(t), \label{QLEFSCAD2} 
\end{eqnarray}
\end{mathletters}
where $X_{in}(t)=b_{in}^{\dagger}(t)+b_{in}(t)$.

In the cold damping scheme of Ref.~\cite{HEIPRL} feedback is provided 
by the radiation pressure of another laser beam intensity-modulated
by the time derivative of the output homodyne photocurrent, and 
therefore one has the additional term in the mirror QLEs
\begin{equation}\label{DLORO}
\dot{{\cal O}}_{fb}(t)=
\frac{i}{\eta \sqrt{\gamma_{c}}}
\frac{dY_{out}(t-\tau)}{dt}
\left[g_{cd}Q(t),{\cal O}(t)\right].
\end{equation}
Again, in the zero delay limit $\tau =0$, and adiabatically 
eliminating the cavity mode, one has
\begin{mathletters}
\label{QLEFCDAD}
\begin{eqnarray}
\dot{Q}(t) &=& \omega_m P(t) 
 \label{QLEFCDAD1}\\
\dot{P}(t) &=& -\omega_{m} Q(t) -  
  {\gamma_m} P(t)   +\frac{2G\beta}{\sqrt{\gamma_{c}}}
  X_{in}(t) + {\cal W}(t) +f(t) \nonumber \\ 
  &-& \frac{4G\beta 
 g_{cd}}{\gamma_{c}}\dot{Q}(t)-\frac{g_{cd}}{\sqrt{\gamma_{c}}}\dot{Y}_{in}(t)
 + \frac{g_{cd}}{2\sqrt{\gamma_{c}\eta}}\dot{Y}_{in}^{\eta}(t).
\label{QLEFCDAD2}
\end{eqnarray}
\end{mathletters}
We have introduced the new input noise $\dot{Y}_{in}(t)$, with the 
correlation function $\langle \dot{Y}_{in}(t)\dot{Y}_{in}(t')
\rangle = -\ddot{\delta}(t-t')$, and the same holds for the 
associated generalized input noise $\dot{Y}_{in}^{\eta}(t)$.

The two sets of equations (\ref{QLEFSCAD})
and (\ref{QLEFCDAD}) show that the two feedback 
schemes are not exactly equivalent. However it is possible to 
see that they have very 
similar physical effects on the mirror dynamics considering, for 
example, the 
Fourier transform of the mechanical susceptibility in the two cases, 
that is, 
$
\chi_{sc}(\omega)=\omega_{m}/\left[\omega_{m}^{2}+
	g_{1}\gamma_{m}-\omega^{2}+i\omega\left(\gamma_{m}+g_{1}\right)\right]
	$
in the ``stochastic cooling'' feedback scheme of Ref.~\cite{MVTPRL} 
($g_{1}=-4G\beta g_{sc}$), and
$
	\chi_{cd}(\omega)=\omega_{m}/\left[\omega_{m}^{2}
	-\omega^{2}+i\omega\left(\gamma_{m}+g_{2}\right)\right]
	$
in the cold damping feedback scheme of Ref.~\cite{HEIPRL} 
($g_{2}=4G\beta g_{cd}\omega_{m}/\gamma_{c}$).
These expressions show that in both schemes the main effect of 
feedback is the modification of mechanical damping $\gamma_{m} 
\rightarrow \gamma_{m}+g_{i}$. In the stochastic cooling scheme one 
has also a frequency renormalization $\omega_{m}^{2} \rightarrow  
\omega_{m}^{2}+g_{1}\gamma_{m}$, which is however usually negligible 
since $\gamma_{m} \ll \omega_{m}$.
If the gains $g_{i}$ are appropriately chosen, one has a significant 
increase of damping and this increase is at the basis of the cooling mechanism 
proposed in Ref.~\cite{MVTPRL} and realized in Ref.~\cite{HEIPRL}. In 
fact, the mechanical susceptibility at resonance is inversely 
proportional to the damping coefficient and, in the presence of 
feedback, the oscillator becomes much less sensitive to the thermal 
noise, yielding a complete suppression of the resonance 
peak in the noise power spectrum. 

The classical treatment of Ref.~\cite{HEIPRL} 
(see also \cite{hei2}) is equivalent
to replace the QLEs (\ref{QLEFSCAD})
and (\ref{QLEFCDAD}) with classical stochastic equations in which the 
back action noise $2G\beta X_{in}(t)/\sqrt{\gamma_{c}}$ and the 
feedback-induced noise terms proportional to $Y_{in}(t)$,
$Y_{in}^{\eta}(t)$, $\dot{Y}_{in}(t)$ and 
$\dot{Y}_{in}^{\eta}(t)$ are neglected. As a consequence,  
within the classical approach, the only effect of 
feedback is damping renormalization $\gamma_{m} 
\rightarrow \gamma_{m}+g_{i}$, while the increase of noise for 
increasing feedback gain, due to the presence of the feedback-induced 
terms, is completely missed. This gives the wrong impression
that, at least in principle, 
unlimited cooling can be achieved for increasing feedback gain. 
With a quantum treatment, the tradeoff 
between damping renormalization and feedback-induced noise leads to the 
existence of an {\em optimal feedback gain}, corresponding to the best 
achievable cooling. This limit on thermal noise suppression 
would be present even in the case of an 
ideal feedback loop with no electronic noise. It
is a manifestation of quantum effects, showing the conceptual difference
with a purely classical description of cooling.

When the classical force we want to detect is characterized by a 
characteristic frequency, say,
  $
f(t)=f_{0}\exp\left[-(t-t_{1})^{2}/2\sigma^{2}\right]
  	\cos\left(\omega_{f}t\right)
  	$,
spectral measurements are commonly performed and the detected signal is 
\begin{equation}
  	S(\omega)= \left|\int_{-\infty}^{+\infty}dt e^{-i\omega t}\langle 
  	Y_{out}(t)\rangle F_{T_{m}}(t)\right|,
  	\label{signal}
  \end{equation}
where $F_{T_{m}}(t)$ is a ``filter'' function, approximately 
equal to one in the time interval $[0,T_{m}]$ in which the spectral 
measurement is performed, and equal to zero otherwise. Using 
Eq.~(\ref{adiab}) and the input-output relation, 
the signal can be rewritten as
\begin{equation}
  	S(\omega)= \frac{8G\beta \eta}{2\pi \sqrt{\gamma_{c}}}
  	\left|\int_{-\infty}^{+\infty}d\omega' \chi(\omega')
  	\tilde{f}(\omega') \tilde{F}_{T_{m}}(\omega -\omega')\right|,
  	\label{signal2}
  \end{equation} 
where $\tilde{f}(\omega)$ and $\tilde{F}_{T_{m}}(\omega )$ 
are the Fourier transforms of the force and 
of the filter function, respectively, and $\chi(\omega)$ is equal to
$\chi_{sc}(\omega)$ or $\chi_{cd}(\omega)$, according to the 
feedback scheme considered. The noise associated with the measurement
of the signal of Eq.~(\ref{signal}) is given by
\begin{eqnarray}
  	 N(\omega)&=& \left\{\int_{-\infty}^{+\infty} dt F_{T_{m}}(t) 
  	\int_{-\infty}^{+\infty} dt' F_{T_{m}}(t') e^{-i\omega (t-t')} \right.
  	\nonumber \\
  && \left. \times \langle Y_{out}(t)Y_{out}(t')\rangle_{f=0} \right\}^{1/2},
  	\label{noise}
  \end{eqnarray}
where the subscript $f=0$ means evaluation
in the absence of the external force. Using again 
Eq.~(\ref{adiab}) and the input noises correlation functions, the 
spectral noise can be rewritten as
\begin{eqnarray}
  && N(\omega)= \left\{ \frac{(8G\beta \eta)^{2}}{\gamma_{c}}
  	\int_{-\infty}^{+\infty}dt F_{T_{m}}(t) 
  	\int_{-\infty}^{+\infty}dt' F_{T_{m}}(t') \right. \nonumber \\
  	&&\left. \times e^{-i\omega 
  	(t-t')}C(t,t')+\eta \int_{-\infty}^{+\infty}dt F_{T_{m}}(t)^{2} 
  	\right\}^{1/2},
  	\label{noise2}
  \end{eqnarray}
where $C(t,t')=\langle Q(t)Q(t')+Q(t')Q(t)\rangle/2 $ is the 
symmetrized correlation function of the oscillator position.
Spectral measurements are usually performed in the stationary case, 
that is, using a measurement time $T_{m}$ much 
larger than the oscillator relaxation time, $T_{m}\gg 1/\gamma_{m}$.
In this limit one has
$F_{T_{m}}(t) \simeq 1$, $\forall t$, and the signal $S(\omega)$ 
simply becomes
$S(\omega)= 8G\beta \eta \left| \chi(\omega)\tilde{f}(\omega) 
\right|/2\pi \sqrt{\gamma_{c}}$.
The oscillator in this case is relaxed to equilibrium and 
$C(t,t')$ in Eq.~(\ref{noise2}) is replaced by the {\em stationary} 
correlation function $C(t-t')$. Defining the 
measurement time $T_{m}$ so that $T_{m}=\int dt F_{T_{m}}(t)^{2}$,  
Eq.~(\ref{noise2}) assumes the usual form
\begin{equation}
  	N(\omega)= \left\{ \left[\frac{(8G\beta \eta)^{2}}{\gamma_{c}}
  	N_{Q}(\omega)+\eta\right]T_{m}  
  	\right\}^{1/2},
  	\label{noisesta}
  \end{equation} 
where $N_{Q}(\omega)=\int dt e^{-i\omega \tau}C(\tau) $. In this 
stationary case, the SNR can be calculated for both 
feedback schemes, and one gets
\begin{eqnarray}
  	&&\frac{S(\omega)}{N(\omega)}_{st}= 
  	|\tilde{f}(\omega)|\left\{T_{m}\left[\frac{\gamma_{m}
  	\omega}{2\omega_{m}}\coth\left(\frac{\hbar \omega}{2kT}\right)+
  	\frac{4G^{2}\beta^{2}}{\gamma_{c}} \right. \right. \nonumber \\
  	&& \left. \left. +\frac{\gamma_{c}}{64
  	G^{2}\beta^{2}\eta}\left(\frac{g_{2}^{2}
  \omega^{2}}{\omega_{m}^{2}} 
  +\frac{1}{\left|\chi_{cd}(\omega)\right|^{2}}\right)
  \right]\right\}^{-1/2}
  	\label{snrcd}
  \end{eqnarray}  
for the cold damping scheme, and a similar expression for the 
stochastic cooling scheme ($g_{2}^2\omega^{2}$ is replaced by $g_{1}^2
(\omega^{2}+\gamma_{m}^{2})$ and $\chi_{cd}$ by $\chi_{sc}$).
In both cases,
feedback does not improve this stationary SNR at any frequency,
due to the $g_{i}^{2}$ term.
This is not surprising because the effect of feedback is to decrease 
the mechanical susceptibility at resonance, so that the oscillator is 
less sensitive not only to the noise but also to the signal.

However, in the case of an {\em impulsive} force with a time duration 
$\sigma \ll 1/\gamma_{m}$, the force spectrum could still be
well reproduced even if a much 
smaller time $T_{m}$, such that $\sigma \ll T_{m} \ll 1/\gamma_{m}$, 
is used. This corresponds to a nonstationary situation because
the system is far from equilibrium during the whole 
measurement. In this case, the noise spectrum is very different from the 
stationary form of Eq.~(\ref{noisesta}) and it is mostly determined by the 
initial state of the oscillator. It is therefore quite natural
to devise a strategy in which the feedback cooling scheme is applied
before the 
measurement, so that the state at the beginning of the measurement is 
just the cooled, equilibrium state in the presence of feedback, and 
turn off the feedback during the spectral measurement. In this way the 
noise remains small during the whole measurement because the heating 
time $1/\gamma_{m}$ is much larger than $T_{m}$, while at the same 
time the signal is not significantly suppressed because the mechanical 
susceptibility is just that in the absence of feedback. One expects 
that as long as the measurement time is sufficiently small,
$T_{m} \ll 1/\gamma_{m}$, the SNR for the detection of the impulsive 
force (which has now to be evaluated using the most general expressions 
(\ref{signal2}) and (\ref{noise2})) can be significantly increased by 
the above nonstationary strategy.

This scheme can be straightforwardly applied whenever the ``arrival 
time'' $t_{1}$ of the impulsive force
is known: feedback has to be 
turned off just before the arrival of the force. However, the 
scheme can be easily adapted to the case of an impulsive force with an 
unknown arrival time, as it is the case of
a gravitational wave passing through an interferometer. In this case 
it is convenient to repeat the process many times, i.e., 
subject the oscillator to cooling-heating cycles. In fact, cyclic cooling 
has been proposed, in a qualitative way, to cool the violin modes of 
a gravitational waves interferometer in \cite{hei2}. Feedback is 
turned off for a time $T_{m}$ during which the spectral measurement 
is performed and the oscillator starts heating up. Then feedback is 
turned on and the oscillator is cooled, and then the process is 
iterated. Cyclic cooling is efficient if the cooling time $T_{cool}$,
which is of the order of $(\gamma_{m}+g_{i})^{-1}$, is much 
smaller than $T_{m}$. This is verified at sufficiently large gains 
and it has been experimentally proved in \cite{hei2}. 
In the impulsive force limit $\sigma \ll T_{m}$, 
the performance of the scheme is well characterized by 
a time averaged SNR, i.e., 
\begin{equation}
\overline{SNR}(\omega)\simeq \frac{1}{T_{m}+T_{cool}}\int_{0}^{T_{m}}dt_{1}
\frac{S(\omega,t_{1})}{N(\omega)}.
	\label{snrmedio}
\end{equation}
This average SNR can be significantly improved 
by cyclic cooling, as it is shown in Fig.~1, 
where $\overline{SNR}(\omega)$ is plotted (full line)
in the case of the impulsive gaussian
force chosen above: one has an improvement by a 
factor $10$ at resonance with respect to the no feedback case (dashed 
line). Parameter values are  
those of Ref.~\cite{HEIPRL}, except that we have 
considered $T=4$K, and the optimal value of the 
feedback gain $g_{2}\simeq 305\gamma_{m}$ maximizing the SNR. The dotted line 
refers to the classical $\overline{SNR}(\omega)$ calculated neglecting 
all quantum noises: the noise is underestimated, with a $15 \%$ error
in the SNR at resonance.
This shows that quantum noise has an appreciable effect already at 
liquid He temperatures and that a fully quantum treatment is 
needed for a faithful description of the physics. It is possible to 
see that at $T=300$K, at the corresponding optimal feedback gain,
there is instead no appreciable difference between the 
classical and quantum predictions, and that
$\overline{SNR}(\omega_{m})$ becomes $16$ times larger than
that with no feedback.

\begin{figure}[h]
\centerline{\epsfig{figure=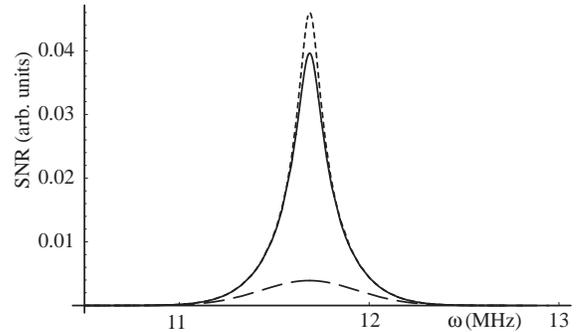,width=7.5cm}}
\caption{\narrowtext{The averaged spectral SNR of Eq.~(\protect\ref{snrmedio})
in the presence of the cold damping feedback scheme with $g_2=82.4$ kHz
in the full quantum treatment (full line), in the classical
approximation (dotted line), and without feedback (dashed line). 
Other parameter values are: $\omega_f=\omega_m =
11.7$ MHz, $\gamma_m=270$ Hz, $G^2 \beta^2/\gamma_c =10.3$ Hz, $T=4$ K,
$\eta=0.99$, $\sigma=3.7$ $\mu$sec, $T_m=11$ $\mu$sec, $T_{cool}=2$ $\mu$sec.}}
\label{pepfig1}
\end{figure}

In conclusion, we have presented the first quantum description of the 
cold damping feedback scheme \cite{coldd}, and we have shown how
the cooling schemes of Refs.~\cite{MVTPRL,HEIPRL}
may be used, within an appropriate nonstationary strategy,
to improve the detection of weak impulsive forces.

\end{multicols}

\end{document}